 \documentclass{emulateapj}

\shorttitle{Star Formation Rate in Local Universe}
\shortauthors{Martin et al.}

\newcommand{\fuvcenter}{1530\AA}

\newcommand{\fuvband}{1350-1750\AA}

\newcommand{\fuvmag}{\ifmmode{FUV}\else{\it FUV~}\fi}
\newcommand{\nuvmag}{\ifmmode{NUV}\else{\it NUV~}\fi}

\begin{document}


\title{The Star Formation Rate Function of the Local Universe}


\author{
Christopher Martin\altaffilmark{1},
Mark Seibert\altaffilmark{1},
Veronique Buat\altaffilmark{4},
Jorge Inglesias-Paramo\altaffilmark{4},
Tom A. Barlow\altaffilmark{1},
Luciana Bianchi\altaffilmark{2},
Yong-Ik Byun\altaffilmark{3}, Jose Donas\altaffilmark{4},
Karl Forster\altaffilmark{1},
Peter G. Friedman\altaffilmark{1},
Timothy M. Heckman\altaffilmark{5},
Patrick N. Jelinsky\altaffilmark{6},
Young-Wook Lee\altaffilmark{3},
Barry F. Madore\altaffilmark{7,8},
Roger F. Malina\altaffilmark{4},
Bruno Milliard\altaffilmark{4},
Patrick F. Morrissey\altaffilmark{1},
Susan G. Neff\altaffilmark{9},
R. Michael Rich\altaffilmark{10},
David Schiminovich\altaffilmark{1},
Oswald H. W. Siegmund\altaffilmark{6},
Todd Small\altaffilmark{1},
Alex S. Szalay\altaffilmark{5},
Barry Y. Welsh\altaffilmark{6}, and
Ted K. Wyder\altaffilmark{1}}

\altaffiltext{1}{California Institute of Technology, MC 405-47, 1200 East
California Boulevard, Pasadena, CA 91125}

\altaffiltext{2}{Center for Astrophysical Sciences, The Johns Hopkins
University, 3400 N. Charles St., Baltimore, MD 21218}

\altaffiltext{3}{Center for Space Astrophysics, Yonsei University, Seoul
120-749, Korea}

\altaffiltext{4}{Laboratoire d'Astrophysique de Marseille, BP 8, Traverse
du Siphon, 13376 Marseille Cedex 12, France}

\altaffiltext{5}{Department of Physics and Astronomy, The Johns Hopkins
University, Homewood Campus, Baltimore, MD 21218}

\altaffiltext{6}{Space Sciences Laboratory, University of California at
Berkeley, 601 Campbell Hall, Berkeley, CA 94720}

\altaffiltext{7}{Observatories of the Carnegie Institution of Washington,
813 Santa Barbara St., Pasadena, CA 91101}

\altaffiltext{8}{NASA/IPAC Extragalactic Database, California Institute
of Technology, Mail Code 100-22, 770 S. Wilson Ave., Pasadena, CA 91125}

\altaffiltext{9}{Laboratory for Astronomy and Solar Physics, NASA Goddard
Space Flight Center, Greenbelt, MD 20771}

\altaffiltext{10}{Department of Physics and Astronomy, University of
California, Los Angeles, CA 90095}


\begin{abstract}

We have derived the bivariate luminosity function for the far ultraviolet (\fuvcenter) and far infrared (60 $\mu$m). We used
matched GALEX and IRAS data, and redshifts from NED and PSC-z.
We have derived a total star formation luminosity function $\phi(L_{tot})$,
with $L_{tot} = L_{FUV}+L_{FIR}$. Using these, we determined 
the cosmic ``star formation rate'' function and density for the local universe.
The total SFR function $\phi(L_{tot})$
is fit very well by a log-normal distribution over five decades of luminosity.
We find that the bivariate luminosity function $\phi(L_{FUV},L_{FIR})$~shows a bimodal behavior, with $L_{FIR}$ tracking
$L_{FUV}$ for $L_{TOT}< 10^{10} L_\odot$, and $L_{FUV}$ saturating at $\sim 10^{10} L_\odot$,
while $L_{TOT}\sim L_{FIR}$ for higher
luminosities.   We also calculate the SFR density
and compare it to other measurements.

\end{abstract}


\keywords{Ultraviolet: galaxies Infrared: galaxies galaxies: fundamental parameters galaxies: luminosity function, mass function galaxies: evolution}


\section{Introduction}

The evolution of the cosmic star formation rate (SFR) density represents
a fundamental constraint on the growth of stellar mass in galaxies over time \citep{madau96,fallpei}.
The distribution of star formation rates, or the ``SFR Function'', 
in galaxies is potentially also a fundamental constraint on cosmological models and on the physics
of star formation in galaxies. 

While a number of SFR metrics have been used in the past, perhaps the most direct measurement of
SFR is the bolometric luminosity of massive stars, usually obtained from the sum of far ultraviolet and
far infrared luminosities. With the launch of GALEX, a large, homogeneous, magnitude limited sample of
UV measurements can be combined with the IRAS FIR sample to generate a true bolometric luminosity function in the local universe.
As discussed in this volume by \cite{buat04}, FUV and FIR selected samples have
quite distinct  far ultraviolet (FUV; \fuvband) to far infrared (FIR; 60$\mu$m) luminosity ratios. However, if the samples are large, homogeneous, 
flux limited, and deep enough in both bands, volume dependence can be removed and the 
fundamental bivariate distribution derived for either sample. To provide the most information, the samples can also be combined
\citep{avni80}.

Our goal in this paper is to generate a bolometric
luminosity function and luminosity density for the bands which sample recent star formation. We do this by
using FUV-selected, FIR-selected, and combined samples to estimate the bivariate luminosity function (BVLF)
in $L_{FUV}$ and $L_{FIR}$,~$\phi(L_{FUV},L_{FIR})$,  using the V$_{max}$ method.  
We then bin this BVLF into a single, total luminosity function (TLF)
$\phi(L_{tot})$, where $L_{tot} = L_{FUV}+L_{FIR}$. We provide some simple parametric fits for the TLF, and estimate the cosmic SFR density.
We discuss whether FUV and FIR selected samples provide consistent measurements of these functions.
We conclude with a brief discussion of the implications. Our cosmology is $\Omega_m=0.3$, $\Omega_\Lambda=0.7$, $H_0=70$ km/s/Mpc.

\section{Samples}

We used two samples to generate the TLF: a far-UV-selected sample (FUVS) and a far-IR-selected sample (FIRS).
The FUVS consists of a primary FUV-selected sample and a IRAS FIR matched co-sample. 
The FIRS consists of a primary FIR-selected sample and a GALEX FUV-selected matched co-sample. 
We measured aperture fluxes for all co-sample matches using optical catalog ellipses.  A small fraction in each
co-sample are formally non-detections--the effect of including these (negligible) is discussed in \S 3.

The FUVS was generated using GALEX All-sky Imaging Survey (AIS) and Medium Imaging Survey (MIS) data
\citep{martin04a, morrissey04}.  The sample consists of objects in GALEX Internal Release 0.2 (consisting of
649 AIS and 94 MIS pointings) with FUV$<$17 that
have a catalog entry in NED. The FUV magnitude limit was selected to insure that, if the galaxy
was not detected in the FIR, an IRAS SCANPI \citep{1988ApJS...68..151H,scanpi} upper limit ($\sim$0.1 Jy) would be
meaningful. NED was used to determine the galaxy redshift and size.
An elliptical aperture based on the galaxy size (usually from RC3)
was used to determine the FUV magnitude, since the patchy nature of
galaxies in the FUV occasionally leads to object shredding by the GALEX pipeline.
We verified that the requirement for a NED entry 
did not compromise completeness or the FUV-selected nature of the sample.
We did this using the GALEX/SDSS-DR2 overlap. Out of 32 FUV$<$17 objects that SDSS classified as
galaxies, 31 objects were in NED (the other was a close degenerate star).

The IRAS 60 and 100 micron data for the FUVS has been compiled from the following
sources - listed in the order of preference. 1. Bright Galaxy Catalog \citep{1989AJ.....98..766S}
2. IRAS Large Optical Catalog \citep{1988ApJS...68...91R}. 3. Faint Source Catalog \citep{1990IRASF.C......0M}. 4. SCANPI (v2.4).
Of the 220 galaxies, 83 had no published IRAS fluxes. Detections $\ge$3
sigma were extracted from SCANPI processing for 37 of the 83. The
remaining 46 are 3 sigma upper limits (all are $>$0.1 Jy) as measured from
the SCANPI processing. With regard to SCANPI, the median coadded scan flux
values are always used. We assume all are extended sources.

The FIRS was generated using PSC-z as the primary catalog \citep{2000MNRAS.317...55S}.
GALEX FUV data from the AIS was available (as of 4/8/2004) for 3938 deg$^2$ of the all-sky PSC-z catalog.
In the overlap area, 991 galaxies appear in the PSC-z catalog, with 878 having valid redshift and FUV data,
for a completeness of 89\%. This small incompleteness is unlikely to affect the results. Two methods were used to
correct for shredding of large galaxies, both using
the APM ellipse parameters \citep{1990MNRAS.243..692M,1990MNRAS.242..318S}, 
which are based on second-moment fitting and are scaled to equal the APM detection isophotal area 
(24 mag arcsec$^{-2}$ for O-plates and 23 mag arcsec$^{-2}$ for E-plates). In the first method, all GALEX catalog objects found 
within the optical APM ellipse were summed
to produce a total magnitude. In the second, an aperture magnitude was obtained within the APM ellipse multiplied by two.  
The latter also provides FUV fluxes for PSC-z objects with no GALEX-detected FUV counterpart (112 out of 878 objects).
As we discuss below, including these non-detections (our default) does not affect the results.

\section{Luminosity Functions\label{sec_lf}}

We calculate the bivariate and total luminosity functions using the $1/V_{max}$ weighting method \citep{schmidt68}. 
V$_{max}$ is calculated  for each
object \citep{willmer97}, and for the FUV and the FIR limits
of the sample. For the FUV limits, we account for the field exposure time and local 
extinction (standard GALEX catalog value). A higher extinction reduces $V_{max}[FUV]$.
The adopted $V_{max}$ depends on the treatment of non-detections. For both FIRS and FUVS, we created samples that excluded the
non-detections and included them. When excluding non-detections,
we use $V_{max} = min(V_{max}[FUV], V_{max}[FIR])$,
since an object can only be detected in both samples if it falls within both volumes.
When including non-detections, all sources are formally included with flux estimates obtained in the identical fashion as true detections.
In this case, $V_{max}$ for the co-sample is formally infinite. For the FUVS, $V_{max} = V_{max}[FUV]$, and for the
the FIRS $V_{max} = V_{max}[FIR]$. Finally, both the BVLF and the TLF are 
obtained by summing $1/V_{max}$ into logarithmic luminosity bins, with $\Delta log L=0.5$.

For simplicity in this preliminary study, luminosities are defined as follows: $L_{FUV} = \nu_{FUV} L_{FUV,\nu}$, where $L_{FUV,\nu}$ is the monochromatic
FUV luminosity; $L_{FIR} = \nu_{60} L_{60,\nu}$, where $L_{60,\nu}$ is the monochromatic luminosity at 60 $\mu$m.
The total luminosity is defined as $L_{TOT} = L_{FUV} + L_{FIR}$, and converted to star formation rate using
SFR[M$_\odot$ yr$^{-1}$]$ = 3.5 \times 10^9 L_{TOT}$ [L$_\odot$] \citep{kennicutt1998}.
More complex relations (e.g., $L_{FIR} = 0.65 \nu_{60} L_{60,\nu} + 0.42 \nu_{100} L_{100,\nu}$)
produce similar results but with more dispersion with respect to the simple functional fits discussed below.
We make no k-corrections, as redshifts are quite low ($z \le 0.04$).

The TLF has been calculated for the FUVS (168 objects including non-detections, 136 excluding them), 
the FIRS (878 objects including non-detections, 766 otherwise), and a combined sample. 
We find that the results do not depend on the the inclusion of non-detections, so all results we present include them.
If all three samples include representative galaxies present in the local universe, the
derived luminosity functions should be consistent within errors from sampling and cosmic variance.
The combined sample is generated following the ``incoherent'' combination method
of \cite{avni80}. The FUVS is obtained in a subset of regions covered by the FIRS.
Thus the FUVS adds information only for objects that have $f_{60}<0.6$, the PSC-z 60 $\mu$m limit. The combined
sample therefore consists of the FIRS sample plus the FUVS ($f_{60}<0.6$) subsample (113), for a total of 991~objects.

\begin{figure*}
\plotone{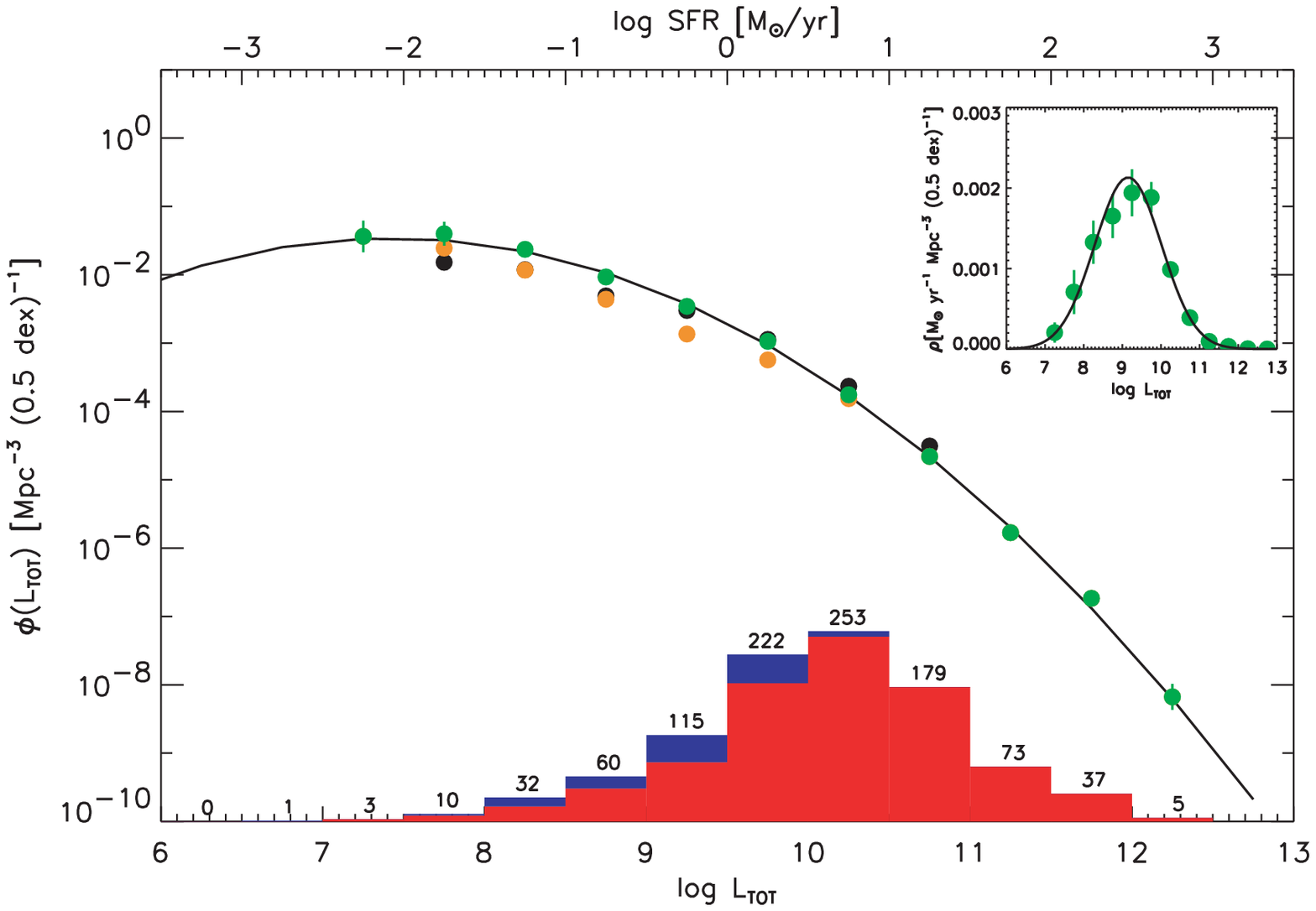}
\caption{The total star formation luminosity function for the local universe. Black points show the FUVS, red points the FIRS,
and green points the combined sample, for the samples that include non-detections. Error bars are shown for the combined
sample.
A histogram shows the number of galaxies in each one-half decade luminosity bin.  The low
luminosity bin with one galaxy has been suppressed.
The curve is the best fit
log-normal function with $\phi_*=0.15$, $\log{L_*}=7.43$, and $\sigma=0.87$. 
The histogram shows the number of FUVS (blue), and FIRS (red) sources in the combined sample in each luminosity bin, with numbers indicating the total in each bin.
INSET: The luminosity density distribution function $L_{TOT}\phi(L_{TOT})$ for the combined sample,
with a line showing the theoretical distribution using the TLF log-normal model fit parameters. 
\label{fig_sfrfunc}}
\end{figure*}

The three TLF are displayed in Figure \ref{fig_sfrfunc}. 
As hoped, the three samples give quite consistent results. The 
FUVS slightly exceeds the FIRS at lower luminosities, while the FIRS fills out the
high luminosity end not represented in the fairly bright cutoff FUVS.  We restrict our analysis here to the combined sample.
The error bars are generated using a bootstrap method. 

As is apparent in Figure \ref{fig_sfrfunc}, a Schechter function provides a very poor fit to the higher luminosity
portion of the TLF, giving a total $\chi^2=115$ for 9 d.o.f. 
On the other hand, a {\it log-normal} function \citep{1990MNRAS.242..318S}
\begin{equation}
\phi(L) d\log{L}= {{\phi_*} \over {\sigma \sqrt{2\pi}}} \exp{  [ {-{(\log{L/L_*})^2} \over {2 {\sigma}^2} } ] }d \log{L}
\end{equation}
provides a remarkably good fit. In this case, the parameters 
$\phi_*=0.150\pm 0.035$, $\log{L_*}=7.43\pm 0.17$, and $\sigma=0.87 \pm 0.03$
(errors generated by the bootstrap) yield a total $\chi^2=8$ for 9 d.o.f. 

The luminosity distribution $L\phi(L)$ is also log-normal, peaking at $L_*^\prime = L_* + \sigma^2 \ln{10}$, or
$L_*^\prime = 9.37$, as we show in the inset of Figure \ref{fig_sfrfunc}.
It can be seen that 50\% of the SFR density comes from galaxies with $\log L_{TOT}<9.4$, or
about 1 $M_\odot/yr$.

\begin{figure*}
\plottwo{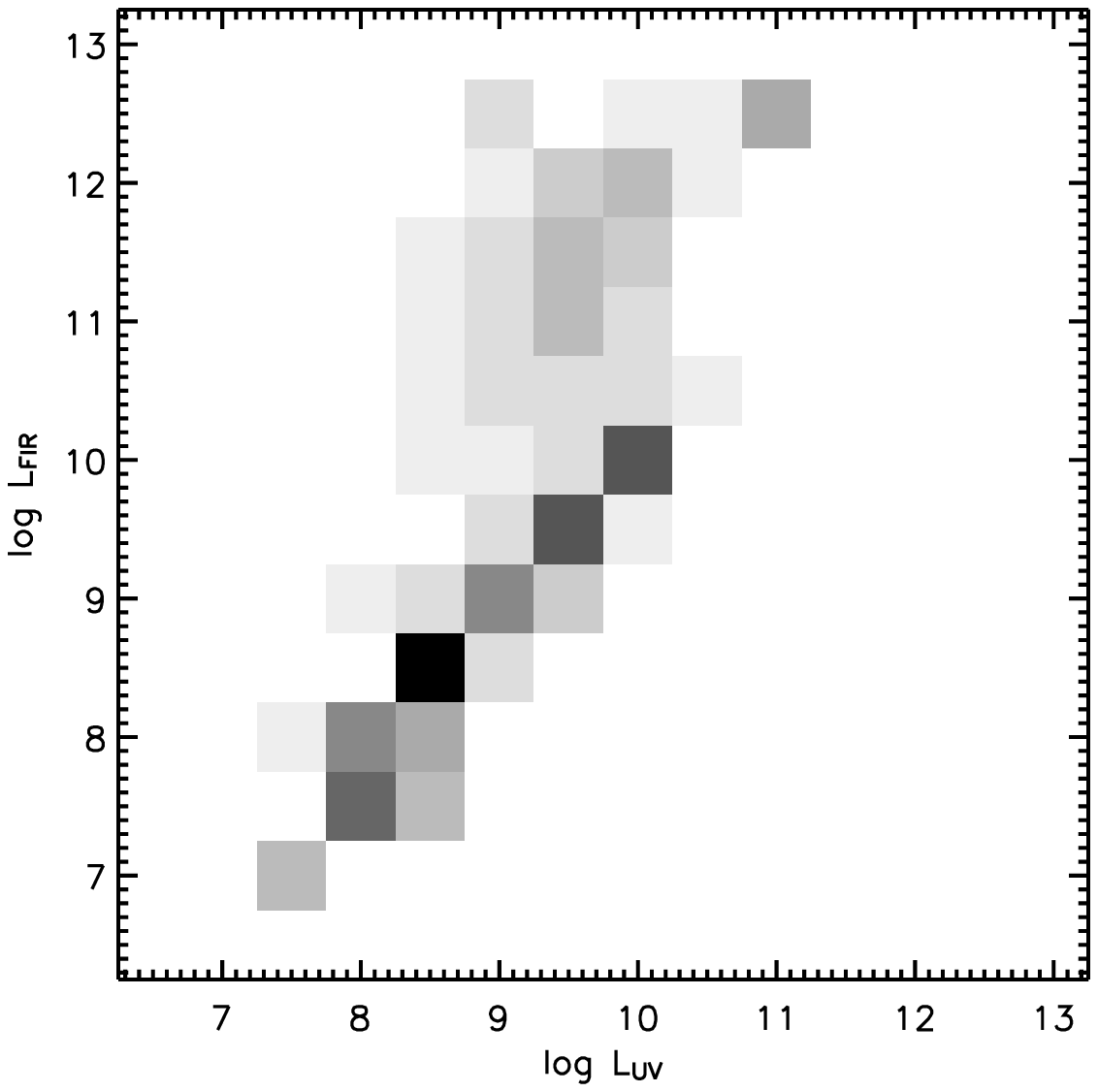}{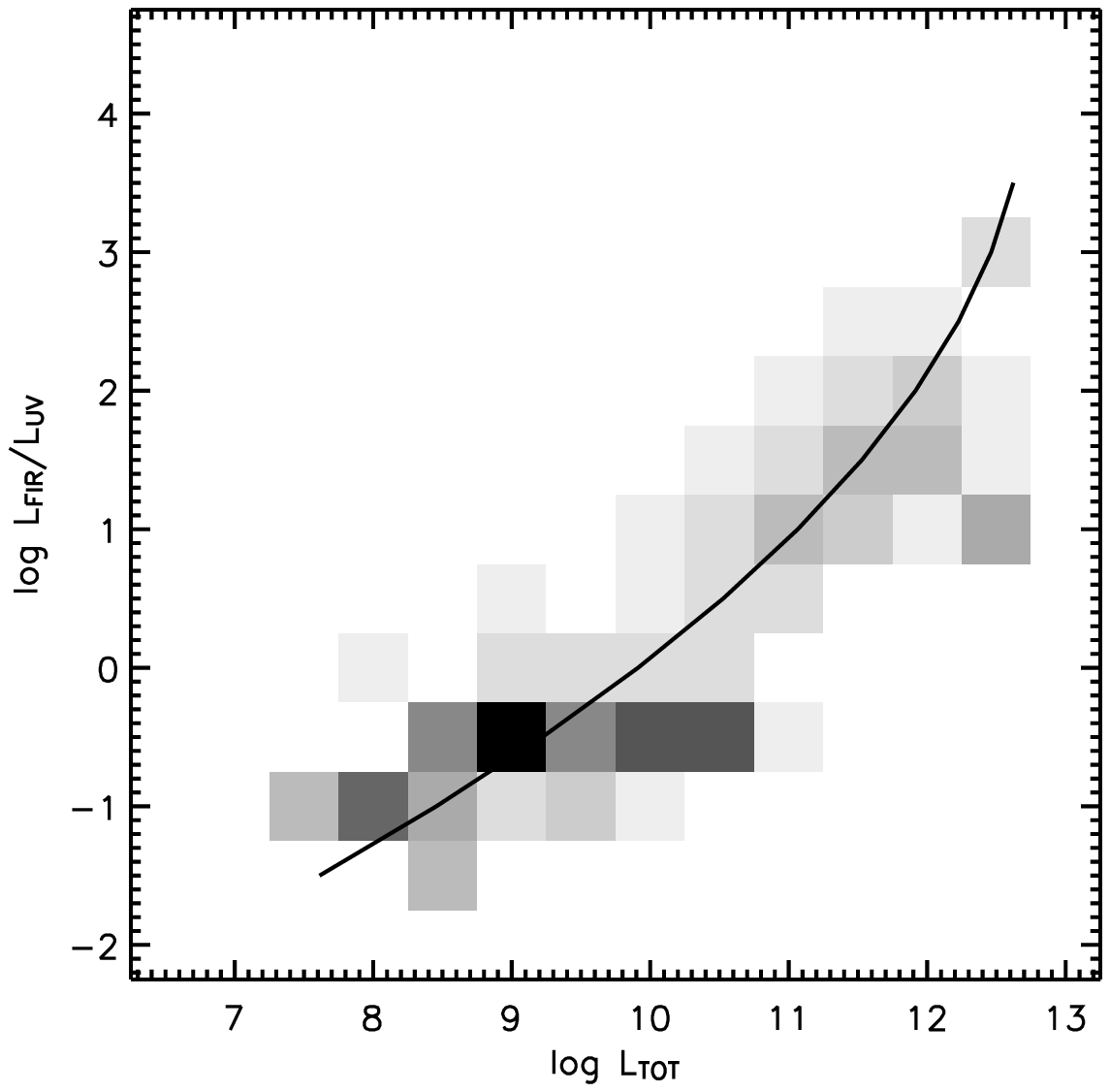}
\caption{LEFT: a) The bivariate luminosity function $\phi(L_{FUV},L_{FIR})$, normalized by $\phi(L_{TOT}$ to
compress the dynamic range. RIGHT: b) The BVLF rebinned as $\phi^\prime(L_{TOT}), L_{FIR}/L_{FUV})$, again
normalized by $\phi(L_{TOT})$. The line shows a quadratic fit discussed in the text.  In both cases the
grayscale density scales linearly with the normalized distribution. \label{fig_bvlf}}
\end{figure*}

We have estimated the BVLF using the combined sample, and we show a 2D histogram normalized by the TLF in Figure \ref{fig_bvlf}a.
The histogram shows quite dramatically why the local FUV LF is well fit by a Schechter function
\citep{wyder04, treyer04}: the FUV luminosity appears to ``saturate'' at $L_{FUV}>10^{10} L_\odot$,
with all increase in the total luminosity coming from FIR radiation. This saturation
apparently occurs at higher redshift, but at a factor of 20 higher $L_{FUV}$ for z=3
\citep{adelberger00}. For $L_{FUV}<10^{10} L_\odot$,
the FIR and FUV luminosities track, but with a slope steeper than unity.
There appears to be a trough between these regions. The trough is statistically significant
as it falls in the range in which the object number distribution peaks.

In Figure \ref{fig_bvlf}b, we show the BVLF rebinned in logarithmic $L_{TOT}$ and $L_{FIR}/L_{FUV}$ bins.
The total luminosity is well correlated with the FIR/UV ratio, and a line is shown with the fit
$L_{TOT}=9.4+1.3(\log{L_{FIR}/L_{FUV}}) - 0.15*(\log{L_{FIR}/L_{FUV}})^2$. The trend of increasing
$L_{FIR}/L_{FUV}$ with increasing $L_{TOT}$ was first noted by \cite{wangheckman96}.

Note also that the FUV projection of the BVLF for the FUVS sample is in excellent agreement
with \cite{wyder04}.

We use the BVLF to calculate the luminosity density using a simple sum:
\newline
$\rho_i = \int \int L_i \phi(L_{FUV},L_{FIR}) d\log{L_{FUV}} d\log{L_{FIR}}$.
From the inset in Figure \ref{fig_sfrfunc}, it is apparent that extrapolation to low or high luminosity
using a model fit to calculate the luminosity density would not alter this result greatly.
The results, using the \cite{kennicutt1998} SFR conversion factor, are
$[L_{FUV}, L_{FIR}, L_{TOT}] = [0.010\pm0.0014, 0.011\pm0.0005, 0.021\pm0.0019] M_\odot yr^{-1} Mpc^{-3}$. Hence
the luminosity density is split roughly 50/50 into primary FUV and reprocessed FIR light.

\section{Discussion}

We have made the first attempt at deriving the bivariate luminosity function for the two bands
which trace the high mass star formation rate in galaxies. The BVLF can be derived from FUV, FIR, or
combined samples. The resulting functions agree for the samples we studied, as hoped. Large, homogeneous,
combined samples that probe the bulk of the BVLF will provide an excellent tool for studying the
relationship between FUV and FIR emission.

We have used the BVLF to generate a total
high mass star formation luminosity function and luminosity density for the local universe.
Our value for the star formation rate density, 0.021 $M_\odot yr^{-1} Mpc^{-3}$ ($<z>\simeq0.02$), is in good agreement with the
estimate of \cite{ha03} using extinction corrected H$\alpha$ of 0.025 $M_\odot yr^{-1} Mpc^{-3}$  and that using
extinction corrected FUV from GALEX \citep{wyder04}, also 0.025 $M_\odot yr^{-1} Mpc^{-3}$.

There are a number of striking features of the BVLF and TLF. We have pointed out the divergence of $L_{FIR}$ and $L_{FUV}$,
behavior similar to that observed when comparing FIR and optical light \citep{1990MNRAS.242..318S,buat98}. 
The BVLF has a bimodal appearance, roughly divided at 3 $M_\odot yr^{-1}$.
Perhaps below the threshold SFR, star formation is an equilibrium process and feedback
is successful at clearing sightlines for FUV emergent flux. With a very high SFR,
the process may take on a non-equilibrium character, where feedback fails to (or has yet to) clear paths for primary radiation.
The FIR/FUV flux ratio appears to be a rough total luminosity proxy.
One explanation is that the highest SFRs occur in the most massive star forming galaxies \citep{brinchmann04}.
The most massive galaxies are the most metal rich \citep{tremonti04}, and a high metallicty ISM
has a high gas to dust ratio. Also, the highest SFRs appear to occur in galaxies with the highest ISM
surface mass densities \citep{kennicutt89}, and higher dust column density.

It has also been known for some time that the FIR luminosity function was not a Schechter function \citep{1990MNRAS.242..318S},
whereas we now know from GALEX that FUV luminosity function is \citep{wyder04}.
Taken together with the diversity of star formation modalities making up our sample, it is therefore surprising
to learn how well the total luminosity function is described by a single log-normal function, over
five decades of luminosity. \cite{norman04} found a log-normal distribution
for the X-ray luminosity function, and notes that it is the expected distribution
for a complex multiplicative random process. Perhaps a unifying physical framework can be found for 
star formation in galaxies ranging from irregular dwarfs to ultra-luminous merging galaxies. 

Optical and near IR band luminosity functions that trace stellar mass are 
well fit by Schechter functions \citep{2003ApJS..149..289B}, as is the UVLF. 
The fact that the TLF is so distinct from that of the fuel, HI \citep{2003AJ....125.2842Z}
and CO \citep{2003ApJ...582..659K}, which are also well fit by Schechter functions, implies that
the star formation efficiency is much higher in luminous star forming galaxies, a widely accepted result.
The fact that the TLF rises far above the stellar mass LF argues strongly that the timescale for star formation
is a falling function of the SFR, perhaps due to fuel consumption or feedback. The BVLF in Figure 3a is clearly a major
clue to the ultimate origin of the high luminosity cutoff of the Schechter function.

With this background, it is extremely interesting to study how the BVLF evolves over cosmic time. 
Lyman break galaxies (LBG) show rest FUV well-fit by Schechter functions, but
with a characteristic UV luminosity a factor of 20 larger. While there is no question that luminous star forming galaxies
were more pervasive in the past, a major question in extragalactic astronomy remains the relationship between
galaxies selected by rest UV vs. rest FIR. Our work suggests that with the launch of GALEX and the Spitzer Space Telescope,
and the flowering of Sub-mm astronomy, a unified approach to
combining rest FUV and FIR information will bring major insights in the next few years.

%comparison to UVLF
%comparison to HI, CO
%comparison to LBGs
%comparison to stellar mass functions -- starburst duration
%evolution of the BVLF
%physical origins

\acknowledgments

GALEX (Galaxy Evolution Explorer) is a NASA Small Explorer, launched in April 2003.
We gratefully acknowledge NASA's support for construction, operation,
and science analysis for the GALEX mission,
developed in corporation with the Centre National d'Etudes Spatiales
of France and the Korean Ministry of 
Science and Technology. The grating, window, and aspheric corrector were supplied by France.
We also acknowledge valuable comments from the referee.

\clearpage

\clearpage

%\clearpage
\clearpage

\end{document}